\documentstyle[11pt,aaspp4]{article}

\newcommand{\ts}{\thinspace}

\begin{document}

\title{THE {\it IRAS} 1-JY SURVEY OF ULTRALUMINOUS INFRARED GALAXIES:\\
       I. The Sample and Luminosity Function\footnote{This work was part of a 
Ph.D. thesis by D.--C. Kim completed in the Department of Physics and 
Astronomy, University of Hawaii, Honolulu, HI}}

\author{D.--C. Kim\footnote{Current address: Infrared Processing and Analysis 
Center, MS 100-22, California Institute of Technology, Pasadena, CA  91125}, and D. B. Sanders}
\affil{Institute for Astronomy, University of Hawaii\\
       2680 Woodlawn Drive, Honolulu, HI  96822}

\begin{abstract}
A complete flux-limited sample of 118 ultraluminous infrared galaxies (ULIGs) 
has been identified from the {\it IRAS} Faint Source Catalog (FSC).
The selection criteria were a 60 micron flux density greater than 1 Jy in a region of the sky 
$\delta > -40^{\circ}, |b| > 30^{\circ}$.  All sources were subsequently reprocessed 
using coadded {\it IRAS} maps in order to obtain the best available flux estimates 
in all four {\it IRAS} wavelength bands. The maximum observed infrared luminosity is 
$L_{\rm ir} = 10^{12.90}\ L_{\sun}$, and the maximum redshift is $z = 0.268$.  
The luminosity function for ULIGs over the decade luminosity range 
$L_{\rm ir} = 10^{12} - 10^{13}\ L_{\sun}$ can be approximated by a power law 
\ $\Phi (L) \propto L^{-2.35}$ $\lbrack {\rm Mpc}^{-3} {\rm mag}^{-1} \rbrack$.  
In the local Universe ($z \lesssim 0.1$), 
the space density of ULIGs appears to be comparable to or slightly larger 
than that of optically selected QSOs at comparable bolometric luminosities. 
A maximum likelihood test suggests strong evolution for our sample;  
assuming density evolution proportional to $(1+z)^\alpha$ we find 
$\alpha = 7.6\pm3.2$.  Examination of the two-point correlation function 
shows a barely significant level of clustering, $\xi (r) = 1.6 \pm 1.2$,  
on size scales $r \sim 22\ h^{-1}$ Mpc.
\end{abstract}


\section{Introduction}

One of the important results of the {\it Infrared Astronomical Satellite}
\footnote{The Infrared Astronomical Satellite was developed and operated 
by the U.S. National Aeronautics and Space Administration (NASA), the 
Netherlands Agency for Aerospace Programs (NIVR), and the U.K. Science 
and Engineering Research Council (SERC).} ({\it IRAS}) all-sky survey was the discovery of 
a significant population of ultraluminous infrared galaxies 
(ULIGs: e.g. Soifer et al. 1986, 1987; Sanders et al. 1987, 1988a) whose infrared 
luminosities\footnote{$L_{\rm ir} \equiv L(8 - 1000 \micron)$. 
Throughout this paper we assume 
$H_{\rm o} = 75\ {\rm km}\ {\rm s}^{-1} {\rm Mpc}^{-1}, q_{\rm o} = 0.0$} 
($L_{\rm ir} \geq 10^{12}\ L_{\sun}$) are equivalent to the 
bolometric luminosities of optically selected quasars (e.g. Schmidt \& Green 1983). 
The origin of ULIGs and the source of their infrared 
power continue to be subjects of intense research.  

The {\it IRAS} 1-Jy survey of ULIGs was begun in early 1991 as an outgrowth of the original  
{\it IRAS} Bright Galaxy Survey (BGS: Soifer et al. 1986, 1987, 1989).  The BGS, which  
included all extragalactic infrared sources brighter than 5.24 Jy at 60{\ts}\micron\ 
at high Galactic Latitude ($\vert b \vert \gtrsim 30^\circ$), had resulted in 
the discovery of 10 ULIGs.  The original intent of the 1-Jy survey was to increase 
this number of identified ULIGs by at least an order of magnitude in order 
to have a substantially larger sample of the nearest and brightest objects that 
could then be targeted for further ground-based study.      
 
The 1-Jy survey began by using the second version of the 
the {\it IRAS Point Source Catalog} (1988: hereafter PSC88).   
However, the PSC88 was abandoned in late 1992 following the release of the 
{\it IRAS Faint Source Catalog - Version 2} (Moshir et al. 1992: hereafter FSC92).  
The FSC92 provided more accurate fluxes\footnote{The FSC92 incorporates the final 
calibration scale (c.f. Pass-3 calibration) that removed inconsistencies in the previous 
point source fluxes computed prior to 1991 versus those computed using 
the more accurate {\it IRAS} data products such as ADDSCAN/SCANPI (Helou et al. 
1988).  Pass-3 calibration had the largest impact on the IRAS 60{\ts}\micron\ and 
100{\ts}\micron\ bands where typical flux changes, compared to values in the PSC88,  
were $\sim${\ts}5--10{\ts}\% with a maximum change of $\sim${\ts}15{\ts}\% for 
the sources in our sample.} and positions than the PSC88 and is now considered 
the standard reference catalog for {\it IRAS} sources.  Section 2 provides a 
complete description of the selection criteria used in the 1-Jy survey and describes 
how candidates were selected from the FSC92.  Section 3 describes the redshift 
survey that was subsequently carried out to identify ULIGs.  The final 1-Jy sample    
(including finder charts for each source) is presented at the end of Section 3.  
Section 4 derives the luminosity function for ULIGs, while Section 5 discusses 
the general properties of ULIGs and compares the luminosity function of ULIGs 
with other classes of extragalactic objects. 

More detailed analyses of individual sources in the 1-Jy sample are being presented 
in companion papers. Optical spectra of ULIGs that were obtained by Kim (1995) 
are analyzed in Kim, Veilleux \& Sanders (1997: Paper II).  More recently 
obtained optical spectra for the entire 1-Jy sample are analyzed in Veilleux, Kim, 
\& Sanders (1998).   Near-infrared spectra for ULIGs that were 
obtained with the United Kingdom Infrared Telescope (UKIRT) on Mauna Kea are 
presented in Veilleux, Sanders, \& Kim (1997; 1998).  Optical and near-infrared 
images that were obtained by Kim (1995) for the majority of the ULIGs 
are analyzed in Kim, Sanders, \& Mazzarella (1998: Paper III).

\section{The {\it IRAS} FSC Survey}

\subsection{Selection Criteria}

Four selection criteria were used to extract candidate ULIG sources from the 
FSC92. Criteria 1--3 follow the general selection methods 
adopted earlier for the {\it IRAS} BGS.  An important fourth criteria was added following 
the detailed analysis of the properties of {\it IRAS} galaxies in the BGS by Soifer \& 
Neugebauer (1991).  The four criteria are:  

\vskip 0.05in

\noindent
{\it (1) All sources were required to have $f_{60} > 1$~Jy\ \footnote{\ The quantities $f_{12}$, 
$f_{25}$, $f_{60}$, and $f_{100}$ represent the {\it IRAS} flux densities in Jy 
at 12{\ts}\micron, 25{\ts}\micron, 60{\ts}\micron, and 100{\ts}\micron\ 
respectively.}, with a ``high'' or ``moderate'' flux quality index (Moshir et al. 1992)}\ \  
This limit is $\sim${\ts}4{\ts}$\sigma$ above the typical threshold at 60{\ts}\micron\ for 
inclusion in the FSC.  

\noindent
{\it (2)} As in the original BGS, {\it all sources were required to be at high 
Galactic latitude ($\vert b \vert > 30^\circ$), and to be easily accessible for 
redshift measurements (in this case, $\delta > -40^\circ$ to allow for observing 
from Mauna Kea).}\ \   
The Galactic latitude restriction was imposed 
to avoid serious contamination by infrared emission from the Galactic plane. 
The total sky coverage represented by the current survey, excluding small regions 
of the sky not surveyed by {\it IRAS} or with less than two {\it IRAS} visits (see 
{\it IRAS Catalogs and Atlases Explanatory Supplement}) 
is $\sim${\ts}16,300{\ts}sq. deg.  Compared 
to the original BGS, the current survey area is approximately 10\% larger
due primarily to the lower declination limit used at Mauna Kea. 

\noindent
{\it (3)} As in the original BGS, {\it all sources were required to have 
$f_{60} > f_{12}$ in order to exclude infrared-bright stars which are known 
to peak around 12{\ts}\micron\  (e.g. Cohen et al. 1987).}\ \  This criteria 
has been shown to be a very efficient means of selecting for galaxies 
while weeding out most Galactic objects.  Our experience with the BGS 
showed that the remaining sources of confusion are from relatively 
nearby planetary nebulae, heavily reddened stars, and reflection nebulae. 
All galaxies in the BGS have $f_{60} > f_{12}$, and this requirement is 
particularly robust in the case of ULIGs where $f_{60}/f_{12}$ 
is typically $>${\ts}20 (see Soifer \& Neugebauer 1991).  

\noindent
{\it (4) All sources were also required to have ``warm'' $f_{60}/f_{100}$ 
colors:\ log $(f_{60} / f_{100}) > -0.3$.}\ \   Soifer \& Neugebauer 
(1991) showed that for galaxies in the BGS, the mean $f_{60}/f_{100}$ flux ratio 
increases systematically with increasing infrared luminosity, and at the 
highest observed luminosities (i.e. for the 10 ULIGs in the BGS) the mean 
ratio is $-0.11$ with a range of $-0.2$ to +0.13.   
Subsequent analysis of an additional 25 ULIGs from both the 2-Jy 
Survey of Strauss et al. (1992) and the BGS-Part{\ts}II (Sanders et al. 1995) 
revealed nearly the same mean and range of far-infrared 
colors.  Therefore, it was decided that the criterion of 
log ($f_{60}/f_{100}) > -0.3$ would not exclude legitimate ULIGs, and 
that it could indeed be used to exclude a large fraction of the more numerous lower-luminosity 
galaxies, the vast majority of which have cooler $f_{60}/f_{100}$ colors.

\subsection{Selection of ULIG Candidates}

Our search of the FSC92 using the four criteria listed above produced a list of 1630 
candidates from a total of $\sim${\ts}175,000 catalog entries. 
Two steps were then used to further reduce this number; first by cross-correlating 
our list with catalogs of known Galactic and extragalactic sources, and second by 
using a size criteria to eliminate relatively nearby lower luminosity galaxies. 

All 1630 {\it IRAS} sources were cross-correlated with previously published 
astronomical catalogs by using the catalogs available in the NASA/IPAC 
Extragalactic Database (NED) and the Strassbourg Data Base (SIMBAD III). This 
resulted in the exclusion of 576 sources that were either Galactic objects 
(mostly highly reddened stars, planetary nebulae, and reflection nebulae), 
or nearby lower luminosity galaxies.  The NED search also picked up 28 previously 
identified ULIGs -- 10 from the final 
version of the BGS (Soifer et al. 1989), and an additional 18 ULIGs from various 
published lists of {\it IRAS} galaxies.  Another 20 ULIGs were then identified from 
a preprint version of the $f_{60} >${\ts}1.936{\ts}Jy survey eventually published 
in Strauss et al. (1992).  Thus, the net result of the cross-correlations was 
a list of 48 previously identified ULIGs and a reduction in the number of 
remaining unidentified ULIG candidates from 1630 to 1006. 

Palomar Optical Sky Survey (POSS) overlays were produced for the remaining 1006 
unidentified candidates, and the optical images were examined using the POSS prints.  
The majority of these objects appeared to be spiral or irregular galaxies as judged 
by their appearance on the POSS E-prints.  Most of these galaxies had 
optical diameters that were much larger than any of the nearest known ULIGs, and 
were therefore assumed to be relatively nearby, less-luminous, infrared galaxies.  

The following simple procedure was developed in order to exclude nearby, 
lower-luminosity galaxies. An optical diameter twice that of Arp 220 (at $z = 0.0185$, 
the nearest ULIG by a factor of $\sim${\ts}2) as measured from the POSS E-plate, was 
used to set a minimum size above which any of the 
remaining 1006 sources was excluded from the final list of ULIG candidates.
This procedure eliminated 837 sources\ \footnote{\ All of these sources were later 
proven to be lower luminosity galaxies by cross-checking with the published 
redshift surveys of Strauss et al. (1992) and Fisher et al. (1995).}.  
However, 18 sources were put back in the unidentified category due to the fact that 
there were two extended objects within the {\it IRAS} position uncertainty 
ellipse, and one of these objects was smaller than our template diameter.  
This resulted in a final list of 187 unidentified candidate ULIGs. 

Overlays were also produced for the 48 previously identified ULIGs in order to verify that 
the more accurate positions from the FSC92 were still spatially consistent with the 
source, all of which had previously been identified using the PSC88.  All sources except for one 
were found to still lie well within the $\pm${\ts}3{\ts}$\sigma$ FSC92 position uncertainty ellipse. 
That one source, IRAS{\ts}F13442+2321 (Leech et al. 1989) was found to be centered on  
a low luminosity spiral galaxy rather than the previously identified higher redshift galaxy 
which was more than 5{\ts}$\sigma$ away from the {\it IRAS} FSC92 position.  This source was dropped 
from our list.

\section{Identification of ULIGs}

The following procedures were used to determine the final list of ULIGs:\  (1){\ts}Optical 
spectra were obtained of all 187 previously unidentified candidates.  In cases were more 
than one obvious source was found within the {\it IRAS} position uncertainty ellipse 
spectra were obtained of each object.\ \ (2){\ts}luminosities were computed using 
measured redshifts from step {\ts}1 and the fluxes in all four {\it IRAS} bands.\ \ 
(3){\ts}optical spectra at higher spectral resolution were obtained of all ULIGs 
identified from step{\ts}2 and all 47 ULIGs identified from the literature (as part 
of a follow-up program to obtain line-ratio diagnostic information for all ULIGs).
The final list of 1-Jy ULIGs is given in Table 1.

\subsection{Redshift Measurements}

Redshifts were initially obtained for all 187 candidate ULIGs and the 47 
previously identified ULIGs using the University of 
Hawaii 2.2-m telescope on Mauna Kea (Kim 1995).  Optical spectra were obtained during 
five separate observing sessions between 1991, August and 1993, March using the 
Faint Object Spectrograph (FOS) at the f/10 Cassegrain focus of the UH 2.2-m 
telescope.  Wavelength coverage was typically $\lambda = 5,000 - 9,000${\ts}\AA\ 
at a resolution of $\sim${\ts}5{\ts}\AA.  Exposure times varied from 180 sec to 
1800 sec depending on the optical brightness of the source. A Zeiss 135R camera and 
long slit (2.5\arcsec$\times$275\arcsec) were used for all observations except 
during February 1993 when a Nikon 180R camera was used.  All observations were 
obtained with either a Tek 1024$\times$1024 or Tek 2048$\times$2048 CCD. The seeing 
was typically $\sim${\ts}0.8\arcsec -- 1.4\arcsec.  Data reduction was performed 
using the TWODSPEC package in IRAF.  All spectra were de-biased by fitting a high 
order polynomial to the bias column, and flat-fielded with a normalized dome-flat.
Wavelength calibrations were performed using an Fe-Ar arc lamp, and flux calibration 
was done by observing spectroscopic standard stars.  The redshifts were determined 
from Gaussian fits of H$\alpha$, [N{\ts}II] and [S{\ts}II] emission lines 
supplemented on occasion with fits of Mg{\ts}Ib and Na{\ts}ID absorption lines.  
The typical uncertainty in the measured velocities is estimated to be 
$\sim \pm$70 km s$^{-1}$ $(1 \sigma)$.

Out of the 187 previously unidentified candidate fields, 175 were found to contain emission 
line galaxies.  Of the remaining 13 objects, 4 were stars, and 9 turned out to be 
empty fields which most likely are infrared cirrus or small Galactic dark clouds.

Six of the 175 fields were found to contain two equally probable galaxies within the 
$\pm${\ts}3{\ts}$\sigma$ position uncertainty ellipse.  In each of the six cases 
both galaxies had nearly identical redshifts ($\Delta z < 0.002$), and high resolution 
imaging (Paper III) shows that all six appear to be strong tidally 
interacting systems; therefore for these six fields it was decided to attribute the {\it IRAS} 
source to the pair and to continue treating the source as a single entry in our list of ULIGs.

\subsection{Infrared Luminosities}

Using our redshifts for each of the 175 newly identified extragalactic sources and 
published redshifts for the 47 previously identified ULIGs, an infrared luminosity,  
$L_{\rm ir}$, was calculated using the expression 
$$L_{\rm ir}=4\pi D_\ell^2 \times F_{\rm ir} \eqno(1)$$
where $D_\ell$ is the luminosity distance for a $q_{\rm o} = 0$ cosmology 
(Weinberg 1972)   
$$D_\ell={c\over{H_0}}\left(z (1 + 0.5 z)\right) \eqno(2)$$
and the infrared flux, $F_{\rm ir}$, in units of 
$\lbrack 10^{-14}\  {\rm W}\ {\rm m}^{-2} \rbrack$ is defined by 
$$F_{\rm ir}=1.8\times\{13.48\times f_{12}+5.16\times f_{25}+2.58\times 
f_{60}+1.0\times f_{100}\}\   
\footnote{The factor of 1.8 in equation (3) was found by Perault (1987) to
provide a good approximation ($\pm${\ts}5\%) to the ratio of the
8--1000{\ts}\micron\  flux ($\equiv F_{\rm ir}$) and the sum
of the flux in all four {\it IRAS} bands (quantity in brackets), for thermal dust
emission in the temperature range $\sim${\ts}25--300{\ts}K and for dust emissivity
laws $\epsilon \propto \lambda^{- \alpha}$ for $\alpha$ in the range
--0.5 to --2.0.  Since many of the high--luminosity galaxies emit a 
significant portion of their infrared luminosity shortward of 40{\ts}\micron, 
this expression provides a significantly better determination of the total 
infrared flux than the more commonly used $F_{\rm fir}$ determined by fitting 
a single temperature dust model to the 60{\ts}\micron\ and 100{\ts}\micron\ fluxes 
(cf. Appendix B of Cataloged Galaxies and Quasars Detected in the {\it IRAS} 
Survey 1985).}.\eqno(3)$$  The {\it IRAS} fluxes were taken directly from the 
FSC92.  A substantial fraction of the sources only had upper limits listed 
for $f_{12}$, and a smaller fraction had upper limits listed for $f_{25}$.

In an attempt to provide the most accurate luminosities possible, we attempted 
to remove 12{\ts}\micron\ and 25{\ts}\micron\  upper limits by manually 
reprocessing the {\it IRAS} data using the ADDSCAN/SCANPI procedure 
(Helou et al. 1988) and visually inspecting the output.  For $\sim${\ts}50{\ts}\%  
of the objects with 25{\ts}\micron\ upper limits, and for $\sim${\ts}20{\ts}\% 
of the objects with 12{\ts}\micron\  
upper limits, we were able to find a signal at the 2--3{\ts}$\sigma$ level; 
for these cases it was decided to use the new values rather than continue to 
use the upper limits in the FSC92.  For those upper limits that could not be 
removed using ADDSCAN/SCANPI we decided to use a template SED (constructed from  
a subsample of nearby ULIGs) to extrapolate the 25{\ts}\micron\ and 12{\ts}\micron\ fluxes. 
The template used was $f_{12} = 0.20{\ts}f_{25}$, and $f_{25} = 0.14{\ts}f_{60}$ for 
objects with $L_{\rm ir} \leq 10^{12.5}{\ts}L_{\sun}$ and  $f_{25} = 0.23{\ts}f_{60}$ for 
objects with $L_{\rm ir} \geq 10^{12.5}{\ts}L_{\sun}$ (see \S{\ts}5.1).  Given these procedures 
for removing upper limits at 12{\ts}\micron\ and 25{\ts}\micron, and 
the fact that even with the upper limits the sum of the 12{\ts}\micron\ and 
25{\ts}\micron\ contribution to $L_{\rm ir}$ is typically $<${\ts}10{\ts}\%, we 
estimate the final uncertainty in $L_{\rm ir}$ due to the 
12{\ts}\micron\ and 25{\ts}\micron\ flux densities to be a few percent, i.e. 
similar to the measurement uncertainty in the 60{\ts}\micron\ and 100{\ts}\micron\ 
flux densities. 
  
Seventy-one of the 175 galaxies proved to have infrared luminosities large enough 
to be classified as ULIGs.  The remaining 98 objects were lower luminosity
emission line galaxies. 

\subsection{Final Adjustments}

As part of a parallel program to obtain broader wavelength coverage plus 
higher spectral resolution data for all of the ULIGs in the 1-Jy sample (Veilleux, 
Kim, \& Sanders 1998), we were eventually able to obtain our own redshifts for all 
of the 47 previously identified ULIGs while also being able to confirm our redshifts 
for the 71 previously unidentified ULIGs.  For all 71 newly identified ULIGs, and 
for most of the objects taken from the literature, our new redshifts were in 
excellent agreement ($\Delta z < .001$) with previous data.  Notable  
differences ($\Delta z > .002$) were found only for six objects, all of which 
were from the subsample of 47 objects taken from the literature.  
However, none of the 47 previously identified ULIGs fell out of the ULIG sample 
when their luminosities were recomputed using our measured redshifts and the new FSC92 
fluxes.  A comparison of our final redshift measurements with redshift measurements 
determined by others is given in Table 2. The final 1-Jy sample contained $71 + 47 = 118$ ULIGs.

\subsection{The Complete {\it IRAS} FSC 1-Jy Sample}

Table 1 lists source parameters for all 118 ULIGs in the final list of objects in the 
1-Jy survey. Column entries are described in the footnotes.  All source parameters 
in columns 1--12 are from the FSC92, except for the estimated 12{\ts}\micron\ and 
25{\ts}\micron\ fluxes (FQ = a in column 12).  In addition to the measured redshift 
(column 13) and computed $L_{\rm ir}$ (column 15) we have also provided the 
quantity $L_{\rm fir}$ (column 14) for comparison with previously published {\it IRAS} 
galaxy samples which used this quantity as the primary measure of ``far-infrared'' 
luminosity. A more complete description of the various methods used in the past 
to represent the {\it IRAS} luminosities is given in Sanders \& Mirabel (1996).  

\subsubsection{Finder Charts}

To aid in the correct identification of the actual {\it IRAS} source for each of the 
objects in Table 1, we have provided small finder charts in Figure 2 showing the 
{\it IRAS} position uncertainty ellipse properly registered on the red filter of the Palomar 
Digital Sky Survey with a marker to identify the ULIG.  The distribution of offsets 
between the actual source and the {\it IRAS} position listed in the FSC92 
(columns 2--5 of Table 1) is typical of the $\sim${\ts}10--15\arcsec\ (1{\ts}$\sigma$) 
uncertainty quoted for the FSC.  All of the sources in the 1-Jy survey are located 
within the $\pm${\ts}3$\sigma$ uncertainty ellipse. 

\section{The Luminosity Function} 

The redshift and luminosity distributions for the 118 ULIGs listed in Table 1 are 
plotted in Figure 3.  The mean and median luminosity values for the sample are 
$\langle L_{\rm ir} \rangle = 10^{12.28 \pm 0.25}{\ts}L_{\sun}$ 
and $\overline{L_{\rm ir}}  = 10^{12.19}{\ts}L_{\sun}$ respectively.  
The mean and median redshift values for the sample are 
$\langle z \rangle = 0.144 \pm 0.043$ and $\overline{z} = 0.145$ respectively.  
The maximum observed luminosity is $(L_{\rm ir})_{\rm max} = 10^{12.90}{\ts}L_{\sun}$, 
and the maximum observed redshift is $z_{\rm max} = 0.268$.  

The infrared luminosity function of ULIGs was calculated using the following expressions (Schmidt 1968)
$$ \Phi = \left({4\pi\over\Omega} \right) \left(\Sigma{1\over {V_{\rm m}}}\right) \eqno(4)$$ and
$$ \sigma_{\phi}= \left({4\pi\over\Omega} \right) \left(\Sigma{1\over V^2_{\rm m}} \right)^{1/2} \eqno(5)$$
where $\Omega$ is the solid angle of the survey region, and $V_{\rm m}$ is the maximum 
volume within which the object would still be included in the survey. 
$V$ and $V_{\rm max}$ were calculated by determining the luminosity distance 
according to the following procedure. The solar motion with respect to the
Local Group\  --- \ $300 \sin \ell \cos b$ [km s$^{-1}$] 
(de Vaucouleurs, de Vaucouleurs, \& Corwin 1976), where 
$\ell$ and $b$ represent Galactic longitude and latitude respectively\  --- \ was 
subtracted and the resulting velocity was then converted to a luminosity 
distance using equation (2).  A correction for the Virgo flow was not 
included since most of our objects are at distances far beyond the influence 
of the Virgo cluster. The monochromatic flux density can then be written as (Weinberg 1972)
$$f_\nu={L_{\nu(1+z)}\over{4\pi{D_\ell}^2/\left(1+z\right)}} \eqno(6)$$
where $L_{\nu(1+z)}$ is the intrinsic luminosity at frequency $\nu (1+z)$ and $D_\ell$
is the luminosity distance.  From equation (6), we can then calculate a maximum luminosity distance
$D_\ell$(max), which is related to the limiting flux density $F_{\rm min}$ by 
$$D_\ell ({\rm max})=D_\ell\left({f\over{f_{\rm min}}}{1+z\over{1+z_{\rm max}}}\mit\Psi\right)^{1/2}\eqno(7)$$
where the {\it K}-correction term $\mit\Psi$ was calculated by a polynomial fit to the
spectral energy distribution between 25{\ts}\micron\ and 100{\ts}\micron.  
The typical value of $\mit\Psi$ in our sample was found to be 0.92--0.94.  
The luminosity function, $\Phi\ \lbrack {\rm Mpc}^{-3}\ {\rm M/2}^{-1} \rbrack$, 
for the 1-Jy sample is tabulated in Table 3.

\section{Discussion}

The luminosity function for ULIGs provides a basis 
for discussing the relative importance of these sources in 
relation to other populations of extragalactic objects.  However, 
before discussing the luminosity function of ULIGs it 
is important 
to understand just how dominant is the far-infrared emission from 
these objects, and why they have until now mostly escaped 
identification in optical surveys.  

\subsection{Spectral Energy Distributions of ULIGs}
 
The {\it IRAS} fluxes from Table 1 have been combined with optical and 
near-infrared fluxes that have recently been obtained
for a substantial fraction of the 1-Jy sample (Kim 1995, Kim et al. 1998), to compute 
the mean SEDs of ULIGs 
as shown in Figure 4.  The mean SED for the entire sample, (as well as the 
SED for each ULIG in the sample),  peaks in the far-infrared near 60{\ts}\micron.  
For all of the ULIGs in our sample, the 
total infrared luminosity, $L_{\rm ir}$, appears to be by far the dominant fraction 
of the bolometric luminosity.  Eventhough the relative optical/infrared 
luminosity ratio increases for objects with $L_{\rm ir} > 10^{12.4}\ L_{\sun}$ 
(see Figure 4), 
$L_{\rm ir}/L_{\rm opt+ir}$ is still $>${\ts}0.9 for all but three of the ULIGs in 
the 1-Jy sample.  Given that the amount of energy outside the infrared band 
is typically $<${\ts}10{\ts}\% 
of $L_{\rm ir}$ in all but a very few of the 1-Jy ULIGs, we have used $L_{\rm ir}$ as 
a good approximation to the bolometric luminosity for the purpose of comparison with 
$L_{\rm bol}$ calculated for other extragalactic objects.  

Compared to modest infrared galaxies like our own 
($L_{\rm ir} \sim 10^{10}{\ts}L_{\sun}$, $L_{\rm ir} 
\sim L_{\rm opt}$), ULIGs have larger infrared luminosities by at least a factor of 
100 while their optical luminosities have only increased by 
factors of $\sim${\ts}2--5.  The 
mean optical and near-infrared luminosities of ULIGs are in fact 
$\sim${\ts}2.5{\ts}$L^*$
 in the K-band\ \footnote{\ For an $L^*$ galaxy $M^*_{\rm B} = -19.7${\ts}mag 
(Schechter 1976), $M^*_{\rm r} = -20.5${\ts}mag assuming a typical $B-r = 0.75$, 
and $M^*_{\rm K^\prime} = -24.2${\ts}mag (Mobasher, Sharples, \& Ellis 1993).}, 
$\sim${\ts}2.7{\ts}$L^*$ in the r-band, and 
$\sim${\ts}2{\ts}$L^*$ in the B-band (see Sanders \& Mirabel 1996).  Given the fact 
that the nearest ULIGs (i.e. those in the RBGS) have mean redshifts 
$\sim${\ts}0.04, none of these objects 
are included in optical catalogs of bright galaxies (e.g. NGC), and 
only four objects in the 1-Jy sample 
are found in the fainter UGC and Zwicky catalogs.  Thus the IRAS ULIGs 
represent an independent sample when compared 
against the well-known optical catalogs of galaxies.  

Further discussion of the SEDs of ULIGs, including a comparison of SED properties 
with object morphology and spectral type, will be given in Papers II and III.  
Here it is simply worth noting that the  
only obvious correlation of the shape of the infrared SED with galaxy 
properties is an apparent increase in the $f_{25}/f_{60}$  ratio with 
increasing infrared luminosity.  A significant fraction ($\sim${\ts}$1/4$) of the 
ULIGs with $L_{\rm ir} > 10^{12.4}\ L_{\sun}$ 
have $f_{25}/f_{60} > 0.2$, similar to values 
that are typically found in infrared selected 
AGN (e.g. Miley et al. 1984; de Grijp et al. 1985; 
Golombeck, Miley \& Neugebauer 1988). 
Paper II discusses the possibility that the increase in the $f_{25}/f_{60}$ ratio 
is in fact due to an increasing proportion of AGN (including a substantial fraction 
of Seyfert{\ts}1s) at the 
highest luminosities in our ULIG sample, a conclusion that 
is supported by new near-infrared 
spectra of a subsample of 25 ULIGs chosen from the 1-Jy 
survey (Veilleux, Sanders, \& Kim 1997).

\subsection{Comparison of the Luminosity Function of ULIGs with that of Other Extragalactic Objects}

Soifer et al. (1987) presented the first meaningful comparison of the 
luminosity function of infrared galaxies with other classes of optically selected extragalactic 
objects by calculating a bolometric luminosity function for both 
infrared and optical samples.  
Figure 5, which is an updated adaptation of a figure from Soifer et al. (1987), 
illustrates the 
differences and similarities between the more well-known optical classes 
of galaxies and the infrared galaxies revealed by {\it IRAS}, and shows how 
infrared galaxies become an increasingly dominant population 
with increasing bolometric luminosity.  Except for the new 1-Jy sample, the 
comparisons illustrated in Figure 5 only reveal what is found for the 
``local Universe'' (i.e. $z \lesssim 0.1$) in that the redshift surveys of 
the optical samples as well as the RBGS 
typically do not extend much above a mean redshift of $\sim${\ts}0.05.
The one exception is for QSOs, but this has been dealt with in Figure 5 by 
showing the local luminosity function for 
QSOs, $\Phi (M_{\rm B}, z=0)$, as derived by 
Schmidt \& Green (1983). 

The luminosity function 
for normal galaxies (i.e. the Schechter function) has its inflection point 
near $L_{\rm ir} \sim 10^{10.3}{\ts}L_{\sun}$ (i.e. $L^*$), 
and falls exponentially at 
$L_{\rm ir} > 10^{11}{\ts}L_{\sun}$.  The luminosity function for 
 optically selected starbursts is remarkably close to the power-law 
defined by the infrared galaxies at 
$L_{\rm ir} < 10^{11}{\ts}L_{\sun}$, but 
starts to fall precipitously at higher luminosities, with 
few identified objects at $L_{\rm ir} > 10^{11.5}{\ts}L_{\sun}$.  
On the other hand, optically selected 
Seyfert galaxies (primarily type-2 Seyferts) are less numerous 
than both infrared galaxies and optically selected starbursts at 
$L_{\rm ir} < 10^{11}{\ts}L_{\sun}$, but are approximately equal to the 
space density of infrared galaxies over the 
decade luminosity range $L_{\rm ir} = 10^{11-12}{\ts}L_{\sun}$.  
At $L_{\rm bol} > 10^{12}{\ts}L_{\sun}$, only QSOs (i.e. type-1 Seyferts 
with $M_{\rm B} < -23$, for $H_{\rm o} = 75${\ts}km{\ts}s$^{-1}${\ts}Mpc$^{-1}$) 
remain in the optical samples, and figure 5 shows that the 
 space density of PG QSOs in the local Universe is $\sim${\ts}1.5 times 
lower\ \footnote{The original comparison of ULIGs and QSOs 
(Soifer et al. 1987) found that the space density of QSOs in the local Universe 
was $\sim${\ts}3 times 
smaller than that of ULIGs in the BGS.  This difference was reduced to a factor of 
$\sim${\ts}2 by Sanders et al. (1989) after better accounting of the infrared 
emission from QSOs, and finally to the factor of 1.5 reported here due 
to using a better approximation to  
the far-UV and soft X-Ray component of QSOs as given in Elvis et al. (1994).}
than that of ULIGs in the RBGS.  However, the relative space densities of optically 
selected QSOs may be equal to or even slightly exceed that of ULIGs if one accepts the 
most recent results of Wisotzki et al. (1998; see also Krishna \& Biermann 1998), 
who claim to find larger numbers of QSOs in the local Universe (compared to 
the number density of PG QSOs) from their new flux-limited 
sample of bright low-redshift QSO and Seyfert{\ts}1 galaxies drawn from the 
Hamburg-ESO survey.

\subsection{Evolution}

Figure 5 also shows the space density of ULIGs from the 1-Jy sample. 
The space density of ULIGs in the 1-Jy sample appears to be $\sim${\ts}2 times 
larger than that computed for ULIGs in the {\it IRAS} RBGS. 
The 1-Jy ULIG survey samples a larger volume of space ($z_{\rm max} \sim 0.3$ 
with a mean ULIG redshift 
of $z \sim 0.15$) than the RBGS ($z_{\rm max} \sim 0.08$ with a mean redshift of 
$z \sim 0.04$ for ULIGs in the RBGS).  We show below that this increase in 
the space density of ULIGs is consistent with 
strong evolution in the ULIG population. 

To quantify the evidence for evolution in the luminosity function of ULIGs
we used the maximum likelihood test of Saunders et al. (1990) including 
refinements by Fisher et al. (1992).  We also assumed pure density evolution  
partly because our sample is not cosmologically distant enough to differentiate between 
different evolutionary models (e.g. pure density or pure luminosity evolution), but also 
because the most recent results from AGN surveys appear to favor pure density evolution over 
pure luminosity evolution models (e.g. Hassinger 1998; K\"ohler \& Wisotzki 1997).
The probability $p_i$ that a galaxy $i$ of given luminosity is located within
a comoving volume $dV$ centered on its observed position $r_i$
with an isotropic comoving density distribution $N(r)$ is
$$p_i=N(r_i)dV_i\Big/\int^{r_{\rm max}}_{r_{\rm min}}N(r)dV. \eqno(8)$$
If we assume that the comoving density distribution $N(r)$ has the form  $N(r)=N(0)(1+z)^\alpha$,
where $N(0)$ is the local space density of ULIGs, then the overall probability 
of measuring the ensemble of a total of $n$ galaxies each with position $r_i$ is
$$P=\prod_{i=1}^{n}p_i=\prod_{i=1}^{n}\Big[(1+z_i)^\alpha dV_i\Big/\int_{r_{\rm min}}^{r_{\rm max}}(1+z)^\alpha
dV\Big]\eqno(9)$$

\noindent
For an assumed cosmological model with given object redshifts and luminosities,
a maximum likelihood estimate for $\alpha$ can be found by maximizing the probability 
$P$ with respect to $\alpha$.  In this calculation, the maximum distance 
$r_{\rm max}$ was calculated using equation (7) where a polynomial fit to the spectral 
energy distribution is used to determine the  {\it K}-correction $\mit\Psi$.  
For $q_{\rm o} = 0.0$ and $H_{\rm o}$ = 75 (km s$^{-1}$ Mpc$^{-1}$), 
the formal maximum likelihood estimate for our sample was $\alpha = 7.6\pm3.2$  
This suggests strong evolution in the ULIG population.  However, it should be noted that 
the redshift range over which this estimate applies is relatively small ($z \lesssim 0.27$), 
and the corresponding uncertainty in the exponent is relatively large.  

Our value of $\alpha = 7.6 \pm 3.2$ 
assuming a pure density evolution model can be compared with previous 
estimates of evolution that have been determined from other 
redshift surveys of {\it IRAS} galaxies.  (However, we note that no previous analyses 
have focused only on ULIGs, but instead have considered all infrared galaxies 
regardless of luminosity.)  The most relevant measure of evolution for 
our purposes comes from a recent analysis of the `FSS-$z$' sample 
(Oliver et al. 1996) of sources with $f_{60} > 0.2${\ts}Jy from the {\it IRAS} 
Faint Source Data Base.  Oliver et al. (1995) used $\sim${\ts}1200 sources over the 
redshift range $0.02 < z < 0.3$ for which redshifts were available, and 
found $\alpha = 5.6\pm1.6$.  The redshift range is nearly identical to that of our 
1-Jy sample of ULIGs, but the FSS-$z$ sample, of course, includes all sources 
regardless of luminosity.  Oliver et al. (1995) also provide a rather thorough 
accounting of several other determinations of $\alpha$ in the literature, all of 
which have been made using brighter 60{\ts}\micron\ samples, and all of which,  
when properly corrected for Malmquist bias,  find 
lower values still for $\alpha$, including a suggestion of no evolution, e.g. 
$\alpha = 2\pm3$, for the $f_{60} > 1.2${\ts}Jy sample (Fisher et al. 1992) where 
no lower redshift cutoff was imposed.  This latter case is clearly affected by the    
known local overdensity of infrared sources (e.g. Hacking, Condon, 
\& Houck 1987), which when included in the analysis would mimic negative evolution
(i.e. the value of $\alpha$ increases to 4.2$\pm$2.3 for this same sample just by 
imposing a lower redshift cutoff, $z > 0.02$).
Although these various measures of evolution at first appear 
inconsistent, they are in fact consistent with the idea that strong evolution is 
only found for the most luminous {\it IRAS} galaxies, and then only for 
redshifts beyond the influence of `local' large-scale structure 
(e.g. $z \gtrsim 0.05$).  Similar conclusions can also be drawn from previous 
analyses of {\it IRAS} extragalactic source counts (Hacking et al. 1987, Lonsdale 
\& Hacking 1989, Lonsdale et al. 1990, Gregorich et al. 1995) that show evidence 
for strong evolution only at relatively low flux levels ($f_{60} < 1${\ts}Jy). 

It is intriguing to note that strong evolution has also been 
found for optically selected QSOs. 
Schmidt \& Green (1983) find clear evidence that the strength of evolution depends 
on luminosity, with little or no evidence for evolution in objects less luminous 
than QSOs (i.e. $M_{\rm B} > -23$ corresponding to 
$L_{\rm bol} < 10^{12}{\ts}L_{\sun}$) with evolution rapidly becoming stronger 
for more luminous systems.  Although Schmidt \& Green (1983) discussed their 
sample in terms of pure luminosity evolution, an examination of their data 
assuming pure density evolution shows that the luminosity function evolves as 
$(1+z)^{\sim 5-8}$ for QSOs with redshifts and luminosities similar to what is 
sampled by the 1-Jy survey of ULIGs (i.e. for QSOs with 
$M_{\rm B} = -23$ to $-25$ 
and redshifts in the range $z = 0 - 0.3$).
Unfortunately, there are 
insufficient data on higher luminosity 
ULIGs (i.e. $L_{\rm ir} > 10^{12.9}{\ts}L_{\sun}$, 
corresponding to the 
bolometric luminosity of QSOs with $M_{\rm B} < -25$)  
to be able to determine whether the luminosity function for ULIGs continues to 
evolve more strongly with 
increasing luminosity as does the luminosity function for QSOs.  
Deeper infrared surveys, such as are 
now being carried out with the {\it Infrared Space Observatory (ISO)} and 
in the future with
the {\it Wide-field Infrared Explorer (WIRE)} 
and the {\it Space Infrared Telescope Facility 
(SIRTF)}, when combined, should be able to measure the 
luminosity function of ULIGs out to the highest redshifts 
observed for QSOs, and thus will provide a clearer answer as 
to how well ULIGs mimic the 
luminosity function of QSOs.  

\subsection{Clustering}

We have also investigated the possibility of clustering of ULIGs 
using the two-point correlation function $\xi(r)$ (Peebles 1980).
At a comoving separation $r$, the two-point correlation function is given by
$$ \xi(r)=N_{\rm gg}/N_{\rm gr} - 1,\eqno(10)$$
where $N_{\rm gg}$ and $N_{\rm gr}$ are galaxy-galaxy and galaxy-random pair
respectively at a given comoving separation of $r-\Delta r/2<r<r+\Delta r/2$.
The errors were estimated from Poisson statistics
$$\Delta\xi=\sqrt{(\xi+1)/N_{\rm gr}},\eqno(11)$$
In this calculation, the total number of a random sample which
produces a minimum statistical noise was found to be $\sim${\ts}100{\ts}$\times$ 
the total number of our sample.
The results of the two-point correlation function analysis are plotted in Figure 6.
As seen in Figure 6, most of the data points are consistent with no clustering
except for one marginally consistent data point at 
$r = 22\ h^{-1}$ Mpc ($h \equiv H_{\rm o} / 100$) 
where the correlation is $1.6\pm1.2$.

Recently, Mo \& Fang (1993) have detected strong clustering for quasars
at $10 \leq r \leq 50\ h^{-1}$ Mpc.
They have used about 750 quasars with mean redshift $<z> = 1.5$.
To compare their result with ours,
we have evolved Mo \& Fang's result back from their mean redshift 
to our mean redshift $<z> = 0.145$ using the following three evolution models
for the correlation function as proposed by Iovino \& Shaver (1988):

i) Comoving model -- The correlation amplitude $A$, defined by the expression 
$\xi (r) = A{\ts}r^{-1.8}$, has no dependence on redshift and the 
correlation function remains constant at all the time.
In this case the clusters expand as the same rate as the Hubble expansion such as 
might be expected in $biased$ theories of clustering (e.g. Davis et al. 1985).

ii) Stable model -- The correlation amplitude $A$ has a $(1+z)^3$
dependence due to gravitational growth,
so the correlation function has the form $\xi (r)\propto(1+z)^{-1.2}$.
In this model the clusters keep the same proper distance with respect 
to the Hubble expansion.

iii) Collapsing model -- clusters collapse as fast as the Hubble expansion and
the correlation amplitude $A$ has a redshift dependence of $< (1+z)^{-1.2}$.

For these  calculations we have used $A = 74$
from the results of Mo \& Fang (1993).  The resulting correlation functions 
for each evolution model are plotted in Figure 6.
The collapsing model gives the poorest fit, and the comoving model (dashed lines) 
fits better than the stable model.  This is consistent with Boyle \& Mo (1993) who 
find a similar clustering scale length for the low redshift quasars ($z < 0.2$).
If the correlation function evolves according to the comoving model,
then the above result gives some marginal evidence that both ULIGs and quasars
may be derived from the same parent population.

\section{Summary}

The following summarizes the main results from our {\it IRAS} 1-Jy survey of 
ULIGs selected from the {\it IRAS} FSC:

1)\ A complete flux-limited sample of 118  
ULIGs with $f_{60} \geq 1$ Jy have been identified 
in a region of the sky at $\vert b \vert > 30^\circ$, $\delta > -40^\circ$ 
($\sim${\ts}16,300 sq deg).  The maximum observed infrared 
luminosity is $(L_{\rm ir})_{\rm max} = 10^{12.90}{\ts}L_{\sun}$, 
and the maximum observed 
redshift is $z_{\rm max} = 0.268$.  

2)\ The luminosity function, $\Phi\ \lbrack {\rm Mpc}^{-3}\ {\rm mag}^{-1} \rbrack$,  
for ULIGs can be approximated by a power law with spectral index 
$\alpha = -2.35 \pm 0.30$ over the decade luminosity range 
$L_{\rm ir} = 10^{12}-10^{13}\ L_{\sun}$.  In the ``local Universe'' ($z \lesssim 0.1$) the  
space density of ULIGs appears to be a factor of $\sim${\ts}1.5--2 times larger than the space 
density of optically selected QSOs at comparable bolometric luminosities.  However, if one accepts 
recent reported evidence for an increase in the local space density of QSOs, the 
space density of QSOs may be equal to, or even slightly exceed that of ULIGs in the local Universe.  

3) A maximum likelihood test suggests strong evolution in our sample.
Assuming pure density evolution, $\Phi(z) \propto (1+z)^\alpha$, 
we find $\alpha = 7.6\pm3.2$\ .

4)  Examination of the two-point correlation function for our sample shows 
only marginal evidence for large-scale clustering, and only on size scales 
$r \sim 22\ h^{-1}$ Mpc.

\acknowledgments 

It is a pleasure to thank Karl Fisher for allowing us to use his evolution
code, Yiping Jing, HouJun Mo, and Yu Gao for their clustering code and 
discussion, Michael Strauss for providing a copy of the preliminary 1.2 Jy
redshift list prior to publication, and Yong-Ik Byun for helpful discussions.
We also thank our telescope operators on the UH 2.2{\ts}m telescope, 
F. Cheigh and D. Woodworth, for their assistance in obtaining the optical redshifts.
DCK and DBS were supported in part by NASA grants NAG 5-1741 and NAGW-3938 
and NSF grant AST-8919563. 
This research has made use of the NASA/IPAC Extragalactic Database (NED) 
which is operated by the Jet Propulsion Laboratory, California Institute of 
Technology, under contract with the National Aeronautics and Space 
Administration.  

\clearpage




\clearpage

\vskip -0.2truein 

\centerline{FIGURE CAPTIONS}

\vskip 0.2truein

\figcaption[fig1.eps]{Aitoff projection in Galactic coordinates of all sources in 
the 1-Jy survey.  The solid line represents the boundary of the survey area (see text).} 

\figcaption[fig2.ps]{Finder charts for ULIGs in the 1-Jy Survey.  The {\it IRAS} 
position uncertainty ellipse ($\pm 3\sigma$) in shown in overlay on the digital POSS 
red image (2\arcmin$\times$2\arcmin\ field of view).  
A marker is centered on the 
identified ULIG.  In those six cases with two marked sources, deeper optical and 
near infrared CCD images (Paper III) show that both galaxies appear to be strongly 
interacting (see text).}

\figcaption[fig3.eps]{Redshift and luminosity distributions of ULIGs in the 1-Jy 
survey.}
 
\figcaption[fig4.eps]{Mean spectral energy distributions of ULIGs 
(normalized at $\lambda = 60\micron$).  Arrows represent upper limits (2{\ts}$\sigma$).}

\figcaption[fig5.eps]{Comparison of luminosity functions for different classes of 
extragalactic objects.  References:  {\it IRAS} 1-Jy Survey (this paper), {\it IRAS} RBGS 
(Sanders et al. 1998), Palomar-Green QSOs (Schmidt \& Green 1983), Markarian starbursts and 
Seyferts (Huchra 1977), and normal galaxies and cDs (Schechter 1976).  The thick solid 
curve is a fit to the {\it IRAS} RBGS data points.  The thick dashed line connects the 
{\it IRAS} 1-Jy data points.}

\figcaption[fig6.eps]{Two-point correlation functions of ULIGs. 
The comoving, stable and collapse models evolved back to $z = 0.145$ are 
plotted as dashed, long-dashed and dot-long-dashed lines respectively (see text).}

\clearpage


\begin{references}

\reference{a1} Allen, D. A., Norris, R. P., Meadows, V. S., \& Roche, P. F. 
1991, \mnras, 248, 528
\reference{a2} Arakelyan, M. A., Dibai, E. A., Esipov, V. F., \& Markarian, B. E. 
1971, Astrofizika, 7, 177
\reference{a3} Armus, L., Heckman, T. M., \& Miley, G. K. 1989, \apj, 347, 727
\reference{b1} Baan, W. A. 1989, \apj, 338, 804
\reference{betal} Bottinelli, L., Dennefeld, M., Gouguenheim, L., Martin, J. M., 
Paturel, G., \& LeSqueren, A. M. 1987, in Star Formation in Galaxies, ed. C.J. 
Lonsdale-Persson (Washington, D.C.: U.S. GPO), 597
\reference{b2} Boyle, B. J., \& Mo, H. J. 1993, \mnras, 260, 925
\reference{c1} Catalogued Galaxies and Quasars Observed in the IRAS Survey 1989, 
produced by L. Fullmer \& C. Lonsdale, JPL D1932 Version 2, (Pasadena: JPL) 
\reference{c2} Cohen, M., Schwarts, D. E., Chokshi, A., \& Walker, R. G. 1987, \aj, 93, 1199
\reference{d1} Davis, M., Efstathiou, G., Frenk, C. S., \& White, S. 1985, \apj, 221, 371
\reference{d2} de Grijp, M. H. K., Miley, G. K., Lub, J., \& de Jong, T. 1985, \nat, 314, 240
\reference{d3} de Vaucouleurs, G., de Vaucouleurs, A., \& Corwin, H. G., Jr. 1976, 
Second Reference Catalogue of Bright Galaxies (Austin: University of Texas Press)
\reference{e1} Elvis, M., et al. 1994, \apjs, 95, 1
\reference{f1} Fisher, K. B., Huchra, J. P., Strauss, M. A., Davis, M., 
Yahil, A., \& Schlegel, D. 1995, \apjs, 100, 69
\reference{f2} Fisher, K. B., Strauss, M. A., Davis, M., Yahil, A., \& 
Huchra, J. P. 1992, \apj, 389, 188
\reference{g1} Golombek, G., Miley, G. K., \& Neugebauer, G. 1988, \aj, 95, 26
\reference{g2} Grandi, S. A. 1977, \apj, 215, 446
\reference{gn} Gregorich, D.T., Neugebauer, G., Soifer, B.T., Gunn, J.E., 
Herter, T.L. 1995, \aj, 110, 259
\reference{hch} Hacking, P. B., Condon, J. J., \& Houck, J. R. 1987, \apjl, 316, L15
\reference{h98} Hassinger, G. 1998, Astron. Nachr., 319, 37
\reference{hel} Helou, G., Kahn, I. R., Malek, L., \& Boehmer, L. 1988, \apjs, 68, 151  
\reference{h1} Hill, G. H., Wynn-Williams, C. G., \& Becklin, E. E. 1987, \apjl, 316, L11
\reference{h2} Hill, G. H., Heasley, J. N., Becklin, E. E., \& Wynn-Williams, C. G. 
1988, \aj, 95, 1031
\reference{i1} Iovino, A., \& Shaver, P. A. 1988, \apjl, 330, L13
\reference{i2} IRAS Catalogs and Atlases Explanatory Supplement 1988, 
eds. C. Beichman, G. Neugebauer, 
H. Habing, P. Clegg, T. Chester (Washington, DC: U.S. GPO)
\reference{i3} IRAS Point Source Catalog, 1988, (Washington, DC: U.S. GPO) (PSC88)
\reference{k0} Kim, D.--C. 1995, Ph.D. Thesis, University of Hawaii
\reference{k1} Kim, D.--C., Sanders, D. B., \& Mazzarella, J. M. 1998, 
\apj, in preparation (Paper III)
\reference{k2} Kim, D.--C., Veilleux, S., \& Sanders, D. B. 1997, \apj, submitted 
(Paper II)
\reference{kw97} K\"ohler, T., \& Wisotzki, L. 1997, in Quasar Hosts, eds. 
D.L. Clements, I. P\'erez-Fournon, (Berlin: Springer-Verlag), 254
\reference{kb98} Krishna, G., \& Biermann, P. L. 1998, A\&A, 330, L37
\reference{la1} Lawrence, A., Rowan-Robinson, M., Leech, K., Jones, D. H. P., \& 
Wall, J. V. 1989, \mnras, 240, 329
\reference{l1} Leech, K. J., Penston, M. V., Terlvich, R., Lawrence, A., Rowan-Robinson, 
M., \& Crawford, J. 1989, \mnras, 240, 349 
\reference{lh} Lonsdale, C. J., \& Hacking, P. 1989, \apj, 339, 712
\reference{letal} Lonsdale, C. J., Hacking, P., Conrow, T. P., Rowan-Robinson, M. 1990, \apj, 358, 60
\reference{l3} Low, F. J., Huchra, J. P., Kleinmann, S. G., \& Cutri, R. M. 1988, \apj, 327, L41
\reference{m1} MacKenty, J. W., \& Stockton, A. 1984, \apj, 283, 64
\reference{mm} Melnick, J., \& Mirabel, I. F. 1990, A\&A, 231, L19
\reference{m2} Miley, G., Neugebauer, G., Clegg, P. E., Harris, S., Rowan-Robinson, M., 
Soifer, B. T., \& Young, E. 1984, \apj, 278, L79
\reference{m3} Mirabel, I. F. 1982, \apj, 260, 75
\reference{m4} Mo, H. J., \& Fang, L. Z. 1993, \apj, 410, 493
\reference{mse} Mobasher, B., Sharples, R. M., \& Ellis, R. S. 1993, \mnras, 263, 560 
\reference{m5} Moshir, M., et al. 1992, Explanatory Supplement to the IRAS Faint 
Source Survey, Version 2, JPL D-10015 8/92 (Pasadena: JPL)\ (FSC92)
\reference{o1} Oliver, S. , et al. 1995, in Wide-Field  Spectroscopy and the 
Distant Universe, eds. S.J. Maddox \& A. Aragon-Salamanca, (Singapore: World Scientific), 264  
\reference{o2} Oliver, S., et al. 1996, \mnras, 280, 673
\reference{p1} Peebles, P. J. E. 1980, The Large Scale Structure of the Universe
(Princeton: Princeton Univ. Press)
\reference{p2} Perault, M. 1987, PhD Thesis, University of Paris
\reference{s0} Sanders, D. B., et al. 1987, in Star Formation in Galaxies, ed. C.J. Persson, 
(Washington, DC: U.S. GPO), 411 
\reference{s1} Sanders, D. B., Egami, E., Lipari, S., Mirabel, I. F., 
\& Soifer, B. T. 1995, \aj, 110, 1993
\reference{s1a} Sanders, D. B., Mazzarella, J. M.,  Surace, J. A., Egami, E., 
Soifer, B. T., Kim, D.-C., Lipari, S., \& Mirabel, I. F. 1998, \apjs, in preparation
\reference{sm} Sanders, D. B., \& Mirabel, I. F. 1996, \araa, 34, 749 
\reference{sp} Sanders, D. B., Phinney, E. S., Neugebauer, G., Soifer, B. T., \& 
Matthews, K. 1989, \apj, 347, 29
\reference{s2} Sanders, D. B., Soifer, B. T., Elias, J. H., Madore, B. F., Matthews, K., 
Neugebauer, G., \& Scoville, N. Z. 1988a, \apj, 325, 74
\reference{s3} Sanders, D. B., Soifer, B. T., Elias, J. H., Neugebauer, G., \& Matthews, K. 
1988b, \apjl, 328, L35
\reference{s4} Saunders, W., Rowan-Robinson, M., Lawrence, A., Efststhiou, G., Kaiser, N., 
Ellis, R. S., \& Frenk, C. S. 1990, \mnras, 242, 318
\reference{s5} Schmidt, M. 1963, \nat, 197, 1040
\reference{s6} Schmidt, M. 1968, \apj, 151, 393
\reference{s7} Schmidt, M., \& Green, R. F. 1983, \apj, 269, 352
\reference{s9} Soifer, B. T., Boehmer, L., Neugebauer, G., \& Sanders, D. B. 1989, \aj, 98, 766
\reference{sn} Soifer, B. T., \& Neugebauer, G. 1991, \aj, 101, 354
\reference{s8} Soifer, B. T. Sanders, D. B., Madore, B. F., Neugebauer, G., Danielson, G. E., 
Elias, J. H., Lonsdale, C. J., \& Rice, W. L. 1987, \apj, 320, 238
\reference{s8a} Soifer, B. T. Sanders, D. B., Neugebauer, G., Danielson, G. E., 
 Lonsdale, C. J., Madore, B. F., \& Persson, S. E. 1986, \apjl, 303, L41
\reference{s10} Strauss, M. A., \& Huchra, J. P. 1988, \aj, 95, 1602
\reference{s11} Strauss, M. A., Huchra, J. P., Davis, M., Yahil, A., Fisher, K. B., 
\& Tonry, J. 1992, \apjs, 83, 29
\reference{vks} Veilleux, S. V., Kim, D.-C., \& Sanders, D. B. 1998, in preparation
\reference{vsk1} Veilleux, S. V., Sanders, D. B., \& Kim, D.-C. 1997, \apj, 484, 92 
\reference{vsk2} Veilleux, S. V., Sanders, D. B., \& Kim, D.-C. 1998, in preparation  
\reference{s12} Weinberg, S. 1972, Gravitation and Cosmology 
(NewYork: John Wiley and Sons)
\reference{w96} Wisotzki, L., K\"ohler, T., Groote, D., Reimers, D. 1996, A\&A, 115, 227
\end{references}
\end{document}


\newcommand{\ts}{\thinspace} 
\setcounter{page}{29}
\begin{deluxetable}{rrrrrrrrrrrrrrr}
\footnotesize
\tablenum{1}
\tablewidth{0pt}
\tolerance=10000
\tablecaption{The IRAS 1~Jy Sample of ULIGs}
\tablehead{
\multicolumn{1}{c}{Name} &
\multicolumn{1}{c}{R.A.} &
\multicolumn{1}{c}{Dec.} &
\multicolumn{1}{c}{R.A.} &
\multicolumn{1}{c}{Dec.} &
\multicolumn{1}{c}{$\ell$} &
\multicolumn{1}{c}{$b$} &
\multicolumn{4}{c}{$f_\nu(\lambda)$} &
\multicolumn{1}{c}{FQ} &
\multicolumn{1}{c}{$z$} &
\multicolumn{1}{c}{log $L_{\rm fir}$} &
\multicolumn{1}{c}{log $L_{\rm ir}$} \nl
\multicolumn{1}{c}{} &
\multicolumn{4}{c}{} &
\multicolumn{2}{c}{} &
\multicolumn{1}{c}{12$\mu$m} &
\multicolumn{1}{c}{25$\mu$m} &
\multicolumn{1}{c}{60$\mu$m} &
\multicolumn{1}{c}{100$\mu$m} \nl
\multicolumn{1}{c}{IRAS FSC} &
\multicolumn{2}{c}{(B1950.0)} &
\multicolumn{2}{c}{(J2000.0)} &  
\multicolumn{1}{c}{($^\circ$)} &
\multicolumn{1}{c}{($^\circ$)} &
\multicolumn{1}{c}{(Jy)} &
\multicolumn{1}{c}{(Jy)} &
\multicolumn{1}{c}{(Jy)} &
\multicolumn{1}{c}{(Jy)} &
\multicolumn{1}{c}{} &
\multicolumn{1}{c}{} &
\multicolumn{1}{c}{($L_\odot$)} &
\multicolumn{1}{c}{($L_\odot$)} \nl \nl
\multicolumn{1}{c}{(1)} &
\multicolumn{1}{c}{(2)} &
\multicolumn{1}{c}{(3)} &
\multicolumn{1}{c}{(4)} &
\multicolumn{1}{c}{(5)} &
\multicolumn{1}{c}{(6)} &
\multicolumn{1}{c}{(7)} &
\multicolumn{1}{c}{(8)} &
\multicolumn{1}{c}{(9)} &
\multicolumn{1}{c}{(10)} &
\multicolumn{1}{c}{(11)} &
\multicolumn{1}{c}{(12)} &
\multicolumn{1}{c}{(13)} &
\multicolumn{1}{c}{(14)} &
\multicolumn{1}{c}{(15)} 
}
\startdata
00091$-$0738&00:09:10.0&--07:38:47&00:11:43.4&--07:22:06&95.63&--68.11&0.07&0.22&2.63&2.52&1232&0.118&12.13&12.19\nl
00188$-$0856&00:18:53.1&--08:56:07&00:21:26.0&--08:39:29&100.52&--70.21&0.12&0.37&2.59&3.40&1332&0.128&12.20&12.33\nl
00397$-$1312&00:39:43.6&--13:12:35&00:42:14.6&--12:56:09&113.86&--75.65&0.14&0.33&1.83&1.90&aa32&0.261&12.72&12.90\nl
00456$-$2904&00:45:37.3&--29:05:00&00:48:03.6&--28:48:38&326.65&--88.16&0.08&0.14&2.60&3.38&1232&0.110&12.06&12.12\nl
00482$-$2721&00:48:13.6&--27:21:02&00:50:40.0&--27:04:43&49.02&-89.82&0.10&0.18&1.13&1.84&1132&0.129&11.89&12.00\nl
01004$-$2237&01:00:24.8&--22:37:57&01:02:51.2&--22:21:51&152.09&--84.58&0.11&0.66&2.29&1.79&a332&0.118&12.08&12.24\nl
01166$-$0844&01:16:36.7&--08:45:03&01:19:07.1&--08:29:18&143.57&--70.24&0.07&0.17&1.74&1.42&a232&0.118&11.96&12.03\nl
01199$-$2307&01:19:58.2&--23:07:41&01:22:22.1&--22:52:01&183.31&--81.80&0.11&0.16&1.61&1.37&1132&0.156&12.18&12.26\nl
01298$-$0744&01:29:49.9&--07:44:40&01:32:20.2&--07:29:16&151.08&--68.10&0.12&0.19&2.47&2.08&1a32&0.136&12.24&12.27\nl
01355$-$1814&01:35:33.3&--18:14:34&01:37:58.1&--17:59:20&174.91&--75.91&0.06&0.12&1.40&1.74&1a32&0.192&12.31&12.39\nl
\nl
01494$-$1845&01:49:27.1&--18:45:39&01:51:50.4&--18:30:51&184.30&--73.65&0.08&0.15&1.29&1.85&1132&0.158&12.11&12.23\nl
01569$-$2939&01:56:58.0&--29:39:10&01:59:13.1&--29:24:38&225.60&--74.92&0.11&0.14&1.73&1.51&1232&0.141&12.11&12.15\nl
01572+0009&01:57:16.8&00:09:13&01:59:50.7&00:23:44&156.57&--57.94&0.12&0.54&2.22&2.16&2332&0.163&12.36&12.53\nl
02021$-$2103&02:02:08.0&--21:03:56&02:04:28.6&--20:49:35&196.15&-72.19&0.07&0.30&1.45&1.72&1232&0.116&11.85&12.01\nl
02411+0353&02:41:08.8&03:53:54&02:43:45.6&04:06:35&168.20&--48.64&0.08&0.22&1.37&1.95&1232&0.144&12.05&12.19\nl
02480$-$3745&02:48:03.0&--37:45:07&02:50:02.1&--37:32:45&243.12&--63.02&0.05&0.11&1.25&1.49&1132&0.165&12.11&12.23\nl
03209$-$0806&03:20:57.1&--08:06:53&03:23:23.6&--07:56:16&192.01&--49.26&0.10&0.13&1.00&1.69&1132&0.166&12.08&12.19\nl
03250+1606&03:25:00.6&16:06:37&03:27:49.6&16:16:59&168.70&--32.35&0.10&0.15&1.38&1.77&1132&0.129&11.94&12.06\nl
Z03521+0028&03:52:08.5&00:28:20&03:54:42.8&00:37:07&188.39&--37.97&0.11&0.20&2.52&3.62&1222&0.152&12.36&12.45\nl
04074$-$2801&04:07:27.9&--28:01:30&04:09:30.3&--27:53:41&225.90&--46.39&0.07&0.07&1.33&1.72&1a32&0.153&12.07&12.14\nl
\nl
04103$-$2838&04:10:18.8&--28:38:00&04:12:20.2&--28:30:22&226.93&--45.89&0.08&0.54&1.82&1.71&2332&0.118&11.97&12.15\nl
04313$-$1649&04:31:22.6&--16:49:48&04:33:37.7&--16:43:34&213.59&--37.84&0.07&0.07&1.01&1.10&1232&0.268&12.48&12.55\nl
05020$-$2941&05:02:03.9&--29:41:04&05:04:00.8&--29:36:57&231.51&--35.09&0.06&0.10&1.93&2.06&1232&0.154&12.24&12.28\nl
05024$-$1941&05:02:26.3&--19:41:10&05:04:36.9&--19:37:05&220.10&--32.00&0.15&0.14&1.06&1.34&a232&0.192&12.19&12.43\nl
05156$-$3024&05:15:36.6&--30:24:23&05:17:31.9&--30:21:14&233.25&--32.44&0.08&0.10&1.16&1.40&1232&0.171&12.12&12.20\nl
05189$-$2524&05:18:58.9&--25:24:40&05:21:01.4&--25:21:46&227.89&--30.29&0.73&3.44&13.67&11.36&3332&0.042&11.91&12.07\nl
07599+6508&07:59:55.6&65:08:18&08:04:33.1&64:59:49&151.21&32.09&0.26&0.53&1.69&1.73&3332&0.149&12.16&12.46\nl
08201+2801&08:20:07.5&28:01:27&08:23:10.1&27:51:46&195.29&31.31&0.09&0.15&1.17&1.43&1a32&0.168&12.11&12.23\nl
08474+1813&08:47:28.1&18:13:14&08:50:18.0&18:02:01&208.74&34.10&0.10&0.19&1.28&1.54&1132&0.145&12.01&12.13\nl
08559+1053&08:55:58.9&10:53:01&08:58:41.7&10:41:21&217.80&33.07&0.10&0.19&1.12&1.95&1232&0.148&12.03&12.16\nl
\nl
08572+3915&08:57:12.6&39:15:41&09:00:25.0&39:03:56&183.40&41.00&0.32&1.70&7.43&4.59&3332&0.058&12.01&12.11\nl
08591+5248&08:59:10.1&52:48:30&09:02:45.6&52:36:39&165.42&40.94&0.10&0.16&1.01&1.53&1132&0.158&12.02&12.14\nl
09039+0503&09:03:55.7&05:03:34&09:06:33.4&04:51:29&225.01&32.12&0.07&0.12&1.48&2.06&aa32&0.125&11.95&12.07\nl
09116+0334&09:11:38.4&03:34:22&09:14:14.8&03:21:54&227.68&33.04&0.09&0.14&1.09&1.82&1132&0.146&12.00&12.11\nl
09463+8141&09:46:21.6&81:41:37&09:52:58.2&81:27:33&130.08&33.03&0.07&0.07&1.43&2.29&1132&0.156&12.17&12.29\nl
09539+0857&09:53:56.4&08:57:18&09:56:35.5&08:43:01&228.55&44.83&0.15&0.15&1.44&1.04&1132&0.129&11.99&12.03\nl
10035+2740&10:03:36.1&27:40:20&10:06:26.6&27:25:41&202.70&53.51&0.14&0.17&1.14&1.63&1132&0.165&12.10&12.22\nl
10091+4704&10:09:10.0&47:04:34&10:12:16.6&46:49:44&169.86&53.17&0.06&0.08&1.18&1.55&1132&0.246&12.48&12.67\nl
10190+1322&10:19:01.5&13:22:10&10:21:41.9&13:07:01&227.21&52.35&0.07&0.38&3.33&5.57&1232&0.077&11.90&12.00\nl
10378+1108&10:37:52.0&11:08:48&10:40:30.2&10:53:07&234.38&55.21&0.11&0.24&2.28&1.82&1232&0.136&12.21&12.26\nl
\nl
10485$-$1447&10:48:35.3&--14:47:25&10:51:03.7&--15:03:21&264.59&38.69&0.11&0.25&1.73&1.66&1a32&0.133&12.06&12.17\nl
10494+4424&10:49:29.1&44:24:56&10:52:22.2&44:08:59&169.61&60.68&0.12&0.16&3.53&5.41&1232&0.092&12.07&12.13\nl
10594+3818&10:59:26.5&38:18:45&11:02:13.7&38:02:36&180.52&64.67&0.09&0.15&1.29&1.89&1132&0.158&12.12&12.24\nl
11028+3130&11:02:54.2&31:30:42&11:05:37.8&31:14:29&196.47&66.60&0.09&0.09&1.02&1.44&1a32&0.199&12.23&12.32\nl
11095$-$0238&11:09:29.5&--02:37:59&11:12:02.6&--02:54:18&260.45&51.58&0.06&0.42&3.25&2.53&a332&0.106&12.14&12.20\nl
11119+3257&11:11:56.8&32:57:54&11:14:39.3&32:41:33&192.18&68.36&0.17&0.35&1.59&1.52&3332&0.189&12.36&12.58\nl
11130$-$2659&11:13:03.2&--26:59:48&11:15:30.2&--27:16:10&278.02&30.92&0.09&0.20&1.21&1.24&1232&0.136&11.93&12.05\nl
11180+1623&11:18:06.7&16:23:15&11:20:43.9&16:06:49&235.93&66.27&0.08&0.19&1.19&1.60&1132&0.166&12.12&12.24\nl
11223$-$1244&11:22:18.8&--12:44:41&11:24:50.1&--13:01:11&272.56&44.69&0.07&0.16&1.52&2.26&1132&0.199&12.41&12.59\nl
11387+4116&11:38:43.6&41:16:25&11:41:22.4&40:59:47&164.56&70.00&0.12&0.14&1.02&1.51&a132&0.149&11.97&12.18\nl
\nl
11506+1331&11:50:38.9&13:31:18&11:53:13.2&13:14:36&255.63&70.56&0.10&0.19&2.58&3.32&1a32&0.127&12.20&12.28\nl
11582+3020&11:58:13.1&30:20:58&12:00:46.9&30:04:16&195.13&78.51&0.10&0.15&1.13&1.49&1132&0.223&12.37&12.56\nl
Z11598$-$0112&11:59:52.7&--01:12:31&12:02:26.5&--01:29:14&278.58&59.05&0.14&0.38&2.39&2.63&1122&0.151&12.32&12.43\nl
12018+1941&12:01:51.1&19:41:50&12:04:24.5&19:25:08&246.00&76.74&0.11&0.37&1.76&1.78&1232&0.168&12.29&12.44\nl
12032+1707&12:03:12.5&17:08:01&12:05:45.9&16:51:19&254.78&75.26&0.14&0.25&1.36&1.54&1a32&0.217&12.42&12.57\nl
12072$-$0444&12:07:11.4&--04:44:32&12:09:45.4&--05:01:14&283.98&56.31&0.12&0.54&2.46&2.47&1332&0.129&12.19&12.35\nl
12112+0305&12:11:13.7&03:05:15&12:13:47.3&02:48:34&281.00&64.07&0.12&0.51&8.50&9.98&a332&0.073&12.21&12.28\nl
12127$-$1412&12:12:43.3&--14:12:58&12:15:18.1&--14:29:38&289.94&47.46&0.13&0.24&1.54&1.13&1232&0.133&12.04&12.10\nl
12265+0219&12:26:34.1&02:19:37&12:29:07.6&02:03:02&289.96&64.36&0.55&0.90&2.06&2.89&3332&0.159&12.32&12.73\nl
12359$-$0725&12:35:57.0&--07:25:59&12:38:32.2&--07:42:28&297.35&55.02&0.09&0.15&1.33&1.12&aa32&0.138&11.99&12.11\nl
\nl
12447+3721&12:44:45.5&37:21:52&12:47:09.1&37:05:30&127.86&80.00&0.12&0.10&1.04&0.84&1a32&0.158&12.02&12.06\nl
12540+5708&12:54:05.8&57:08:30&12:56:15.0&56:52:17&121.61&60.24&1.87&8.66&31.99&30.29&3332&0.042&12.32&12.50\nl
13106$-$0922&13:10:37.2&--09:22:13&13:13:14.4&--09:38:06&311.85&52.86&0.12&0.06&1.24&1.89&1122&0.174&12.20&12.32\nl
13218+0552&13:21:50.2&05:52:34&13:24:21.6&05:36:57&324.41&67.10&0.26&0.40&1.17&0.71&3332&0.205&12.38&12.63\nl
13305$-$1739&13:30:33.1&--17:39:38&13:33:15.2&--17:55:01&316.78&43.83&0.09&0.39&1.16&1.04&1232&0.148&11.99&12.21\nl
13335$-$2612&13:33:35.1&--26:12:13&13:36:22.2&--26:27:31&315.28&35.33&0.13&0.14&1.40&2.10&1132&0.125&11.94&12.06\nl
13342+3932&13:34:12.9&39:32:42&13:36:24.5&39:17:25&88.17&74.65&0.06&0.27&1.11&1.63&aa32&0.179&12.17&12.37\nl
13428+5608&13:42:51.4&56:08:15&13:44:41.8&55:53:14&108.11&59.68&0.24&2.28&21.74&21.38&3332&0.037&12.04&12.10\nl
13443+0802&13:44:21.8&08:02:13&13:46:51.4&07:47:15&339.65&66.62&0.12&0.11&1.50&1.99&1132&0.135&12.03&12.15\nl
13451+1232&13:45:06.3&12:32:19&13:47:33.5&12:17:23&347.22&70.17&0.14&0.67&1.92&2.06&a232&0.122&12.03&12.28\nl
\tablebreak
13454$-$2956&13:45:29.0&--29:56:57&13:48:20.2&--30:11:52&317.29&31.08&0.06&0.07&2.16&3.38&1a32&0.129&12.17&12.21\nl
13469+5833&13:46:55.6&58:33:39&13:48:40.1&58:18:46&109.05&57.25&0.05&0.04&1.27&1.73&1a32&0.158&12.11&12.15\nl
13509+0442&13:50:59.1&04:42:55&13:53:30.3&04:28:11&338.82&62.93&0.10&0.23&1.56&2.53&2132&0.136&12.09&12.27\nl
13539+2920&13:53:55.1&29:20:07&13:56:10.9&29:05:29&44.54&75.60&0.09&0.12&1.83&2.73&1232&0.108&11.93&12.00\nl
14053$-$1958&14:05:23.3&--19:58:02&14:08:09.9&--20:12:15&326.36&39.15&0.07&0.14&1.02&1.12&1132&0.161&12.01&12.12\nl
14060+2919&14:06:03.4&29:19:09&14:08:17.5&29:04:57&44.01&72.96&0.10&0.14&1.61&2.42&1232&0.117&11.95&12.03\nl
14070+0525&14:07:00.4&05:25:44&14:09:30.9&05:11:35&346.69&61.24&0.07&0.19&1.45&1.82&1a32&0.265&12.64&12.76\nl
14121$-$0126&14:12:11.0&--01:26:56&14:14:45.8&--01:40:53&341.11&54.89&0.06&0.11&1.39&2.07&aa32&0.151&12.11&12.23\nl
14197+0813&14:19:44.3&08:13:03&14:22:12.6&07:59:24&355.46&61.22&0.17&0.19&1.10&1.66&1132&0.131&11.88&12.00\nl
14202+2615&14:20:16.2&26:15:42&14:22:31.0&26:02:05&35.05&69.64&0.18&0.15&1.49&1.99&a232&0.159&12.18&12.39\nl
\nl
14252$-$1550&14:25:16.2&--15:50:19&14:28:01.4&--16:03:43&334.25&40.86&0.09&0.23&1.15&1.86&1132&0.149&12.04&12.15\nl
14348$-$1447&14:34:52.3&--14:47:22&14:37:37.3&--15:00:20&337.59&40.61&0.07&0.49&6.87&7.07&a332&0.083&12.24&12.28\nl
14394+5332&14:39:28.5&53:32:55&14:41:04.3&53:20:08&92.73&56.88&0.03&0.35&1.95&2.39&a332&0.105&11.91&12.04\nl
14485$-$2434&14:48:30.1&--24:34:14&14:51:24.6&--24:46:33&334.75&30.58&0.11&0.15&1.02&1.05&1132&0.148&11.93&12.04\nl
15001+1433&15:00:11.6&14:33:14&15:02:33.1&14:21:30&16.90&56.93&0.12&0.21&1.87&2.04&1232&0.162&12.28&12.38\nl
15043+5754&15:04:21.6&57:54:46&15:05:38.7&57:43:13&94.69&51.41&0.12&0.07&1.02&1.50&1232&0.151&11.98&12.05\nl
15130$-$1958&15:13:03.5&--19:58:15&15:15:55.6&--20:09:18&343.24&31.18&0.14&0.39&1.92&2.30&1232&0.109&11.93&12.09\nl
15206+3342&15:20:38.5&33:42:15&15:22:38.0&33:31:36&53.45&56.94&0.08&0.35&1.77&1.89&2332&0.125&12.02&12.18\nl
15225+2350&15:22:32.9&23:50:43&15:24:43.8&23:40:11&35.92&55.34&0.07&0.18&1.30&1.48&1332&0.139&11.98&12.10\nl
15327+2340&15:32:46.8&23:40:07&15:34:57.1&23:30:10&36.63&53.03&0.48&7.92&103.33&112.40&3332&0.018&12.12&12.17\nl
\nl
15462$-$0450&15:46:18.0&--04:50:29&15:48:56.6&--04:59:36&2.94&36.44&0.10&0.45&2.92&3.00&a332&0.100&12.03&12.16\nl
16090$-$0139&16:09:05.4&--01:39:25&16:11:41.0&--01:47:06&10.25&33.84&0.09&0.26&3.61&4.87&2322&0.134&12.40&12.49\nl
16156+0146&16:15:36.2&01:46:37&16:18:08.2&01:39:21&14.79&34.39&0.10&0.28&1.13&1.00&1332&0.132&11.88&12.04\nl
16300+1558&16:30:03.7&15:58:08&16:32:19.8&15:51:49&32.43&37.80&0.07&0.07&1.48&1.99&1a32&0.242&12.57&12.63\nl
16333+4630&16:33:22.6&46:30:55&16:34:50.9&46:24:48&72.30&42.43&0.06&0.06&1.19&2.09&1a32&0.191&12.30&12.35\nl
16468+5200&16:46:49.6&52:00:52&16:48:02.2&51:55:40&79.36&39.82&0.06&0.10&1.01&1.04&1232&0.150&11.95&12.02\nl
16474+3430&16:47:24.5&34:30:21&16:49:14.7&34:25:13&56.63&39.12&0.13&0.20&2.27&2.88&1332&0.111&12.02&12.11\nl
16487+5447&16:48:44.1&54:47:38&16:49:47.8&54:42:34&82.90&39.23&0.07&0.20&2.88&3.07&1332&0.104&12.07&12.12\nl
17028+5817&17:02:52.7&58:17:53&17:03:41.8&58:13:48&87.01&36.86&0.06&0.10&2.43&3.91&1332&0.106&12.05&12.10\nl
17044+6720&17:04:29.8&67:20:34&17:04:28.5&67:16:34&97.99&35.12&0.07&0.36&1.28&0.98&1332&0.135&11.97&12.13\nl
\nl
17068+4027&17:06:53.6&40:27:12&17:08:32.0&40:23:25&64.76&36.14&0.08&0.12&1.33&1.41&1232&0.179&12.23&12.30\nl
17179+5444&17:17:54.2&54:44:51&17:18:55.1&54:41:50&82.46&35.04&0.08&0.20&1.36&1.91&1332&0.147&12.07&12.20\nl
20414$-$1651&20:41:28.4&--16:51:08&20:44:17.4&--16:40:14&29.62&--32.27&0.65&0.35&4.36&5.25&1232&0.086&12.06&12.14\nl
21208$-$0519&21:20:51.0&--05:19:53&21:23:28.7&-05:06:59&47.25&-35.91&0.09&0.15&1.17&1.66&1132&0.130&11.89&12.01\nl
21219$-$1757&21:21:55.1&--17:57:37&21:24:42.5&--17:44:40&32.78&--41.65&0.21&0.45&1.07&1.18&3332&0.112&11.69&12.06\nl
21329$-$2346&21:32:54.2&--23:46:03&21:35:45.4&--23:32:36&26.26&--45.91&0.05&0.12&1.65&2.22&aa32&0.125&11.99&12.09\nl
21477+0502&21:47:47.1&05:02:11&21:50:17.6&05:16:14&62.48&--35.57&0.09&0.16&1.14&1.46&1a32&0.171&12.12&12.24\nl
22088$-$1831&22:08:51.5&--18:31:50&22:11:35.5&--18:17:00&37.87&--52.26&0.09&0.07&1.73&1.73&1a32&0.170&12.29&12.31\nl
22206$-$2715&22:20:41.5&--27:15:12&22:23:29.5&--27:00:00&24.49&--57.19&0.10&0.16&1.75&2.33&1132&0.132&12.06&12.19\nl
22491$-$1808&22:49:08.7&--18:08:24&22:51:49.0&--17:52:27&45.18&--60.95&0.05&0.55&5.44&4.45&a232&0.076&12.05&12.09\nl
\nl
22541+0833&22:54:10.4&08:33:12&22:56:41.3&08:49:15&81.20&--44.55&0.09&0.18&1.20&1.48&1132&0.166&12.11&12.23\nl
23060+0505&23:06:01.8&05:05:13&23:08:34.2&05:21:29&81.65&--49.09&0.20&0.43&1.15&0.83&3332&0.173&12.16&12.44\nl
23129+2548&23:12:54.3&25:48:09&23:15:21.6&26:04:31&97.40&--31.99&0.08&0.08&1.81&1.64&1a32&0.179&12.36&12.38\nl
23233+2817&23:23:20.5&28:17:49&23:25:48.7&28:34:19&101.06&-30.64&0.13&0.28&1.26&2.11&1332&0.114&11.83&12.00\nl
23234+0946&23:23:24.0&09:46:22&23:25:56.0&10:02:52&90.89&--47.43&0.06&0.08&1.56&2.11&1a32&0.128&11.99&12.05\nl
23327+2913&23:32:43.0&29:13:29&23:35:12.5&29:30:05&103.67&--30.50&0.06&0.22&2.10&2.81&1232&0.107&11.96&12.06\nl
23389+0300&23:38:57.7&03:00:52&23:41:31.1&03:17:31&91.22&--55.22&0.09&0.35&1.23&1.17&1132&0.145&11.99&12.09\nl
23498+2423&23:49:53.8&24:23:39&23:52:26.4&24:40:20&106.25&--36.28&0.10&0.12&1.02&1.45&1a32&0.212&12.29&12.40\nl
\tablenotetext{}
{{\it Col 1:}\ The {\it IRAS} object name in the Faint Source Database (FSDB).  
The prefix `FSC' is the standard designator for 
sources in the Faint Source Catalog.  The prefix `Z' is used for objects 
that are not in the FSC -- these objects are contained in the Faint Source 
Reject File (see Moshir et al. 1992).}
\tablenotetext{}{{\it Cols 2-5:}\ Right Ascension (hh:mm:ss.s) and Declination (dd:$^\prime\ ^\prime$:$^{\prime\prime}\ ^{\prime\prime}$) of the {\it IRAS} 
source position as listed in the FSDB.  Positions are given for epochs B1950.0 and J2000.0{\ts}.}
\tablenotetext{}{{\it Cols 6-7:}\ Galactic latitude and longitude.  The distribution of all 111 sources in 
Galactic coordinates is given in Figure 1.} 
\tablenotetext{}{{\it Cols 8-11:}\ {\it IRAS} flux densities at 12\micron, 
25\micron, 60\micron, and 100\micron\ as listed in the {\it IRAS} FSDB.} 
\tablenotetext{}{{\it Col 12:}\ Flux quality in each of the four {\it IRAS} bands.  Numbers 1--3 are the 
quality flags adopted for the {\it IRAS} catalogs (1 = upper limit, 
2 = moderate quality, 
3 = high quality).  The letter `a' means that the flux density 
was estimated by us after 
coadding all of the {\it IRAS} data available for the source using 
the ADDSCAN/SCANPI procedure (Helou et al. 1988); 
these `low quality' measurements are typically 2--3{\ts}$\sigma$ detections.}    
\tablenotetext{}{{\it Col 13:}\ Redshifts determined from our high resolution 
optical spectra (see \S4.2).
These redshifts are compared in the Appendix with our previous redshift measurements (e.g. Kim 1995) and with redshift data 
from the literature.}  
\tablenotetext{}{{\it Col 14:}\ Far-infrared luminosity determined using the 60\micron\ and 100\micron\ 
{\it IRAS} fluxes acording to the prescription given in Appendix B 
of Cataloged Galaxies 
and Quasars Detected in the {\it IRAS} Survey (1989) (see also Sanders \& Mirabel 1996).}  
\tablenotetext{}{{\it Col 15:}\ Infrared luminosity computed using the flux in all four {\it IRAS} bands 
according to the expressions given in equations 1--3.}
\enddata
\end{deluxetable}

\clearpage

\begin{deluxetable}{rcclll}
\tablecolumns{6}
\tablenum{2}
\tablewidth{0pt}
\tolerance=10000
\tablecaption{Redshifts}
\tablehead{
\multicolumn{1}{c}{Name} &
\multicolumn{1}{c}{our work} & 
\multicolumn{1}{c}{} &
\multicolumn{2}{c}{previous values\qquad \ \ \ } &  
\multicolumn{1}{c}{} \nl 
\multicolumn{1}{c}{IRAS FSC} &
\multicolumn{1}{c}{$z_{\rm opt}$} &
\multicolumn{1}{c}{} &
\multicolumn{1}{l}{$z_{\rm opt}$} &
\multicolumn{1}{l}{Ref} &
\multicolumn{1}{l}{Other Names}  
}
\startdata
00091$-$0738&0.118&&0.118&(18)\nl
00188$-$0856&0.128&&0.129&(18)\nl
00397$-$1312&0.261&&0.262&(19)\nl
00456$-$2904&0.110&&0.110&(1)\nl
00482$-$2721&0.130&&\nodata&\nodata\nl
01004$-$2237&0.118&&0.118&(1)\nl
01166$-$0844&0.118&&0.118&(19)\nl
01199$-$2307&0.156&&0.156&(1)\nl
01298$-$0744&0.136&&0.136&(18)\nl
01355$-$1814&0.192&&0.191&(19)\nl
\nl
01494$-$1845&0.158&&0.157&(1)\nl
01569$-$2939&0.141&&0.140&(1)&TXS0156-296\nl
01572+0009&0.163&&0.163&(10)&Mrk1014, PG0157+001\nl
02021$-$2103&0.116&&0.116&(1)\nl
02411+0353&0.144&&0.144&(19)\nl
02480$-$3745&0.165&&0.165&(19)\nl
03209$-$0806&0.166&&\nodata&\nodata\nl
03250+1606&0.129&&0.129&(19)\nl
Z03521+0028&0.152&&0.152&(18)\nl
04074$-$2801&0.153&&\nodata&\nodata\nl
\nl
04103$-$2838&0.118&&0.117&(19)\nl
04313$-$1649&0.268&&\nodata&\nodata\nl
05020$-$2941&0.154&&\nodata&\nodata\nl
05024$-$1941&0.192&&\nodata&\nodata\nl
05156$-$3024&0.171&&\nodata&\nodata\nl
05189$-$2524&0.042&&0.043&(13)\nl
07599+6508&0.149&&0.149, 0.148&(14), (19)&[HB89]0759+65\nl
08201+2801&0.168&&\nodata&\nodata\nl
08474+1813&0.145&&0.146&(19)\nl
08559+1053&0.148&&0.149&(8)\nl
\nl
08572+3915&0.058&&0.058&(13)\nl
08591+5248&0.158&&\nodata&\nodata\nl
09039+0503&0.125&&0.125&(19)\nl
09116+0334&0.146&&\nodata&\nodata\nl
09463+8141&0.156&&0.155&(19)\nl
09539+0857&0.129&&\nodata&\nodata\nl
10035+2740&0.165&&\nodata&\nodata\nl
10091+4704&0.246&&0.245&(19)\nl
10190+1322&0.077&&0.077&(18)\nl
10378+1108&0.136&&0.136&(18)\nl
\nl
10485$-$1447&0.134&&0.133&(3)\nl
10494+4424&0.092&&0.092&(18)\nl
10594+3818&0.158&&0.157&(19)\nl
11028+3130&0.199&&\nodata&\nodata\nl
11095$-$0238&0.106&&0.106&(18)\nl
11119+3257&0.189&&0.187&(3)\nl
11130$-$2659&0.136&&\nodata&\nodata&AM1113-270\nl
11180+1623&0.166&&\nodata&\nodata\nl
11223$-$1244&0.199&&\nodata&\nodata\nl
11387+4116&0.149&&\nodata&\nodata\nl
\nl
11506+1331&0.127&&0.127&(18)\nl
11582+3020&0.223&&\nodata&\nodata\nl
Z11598$-$0112&0.151&&0.151&(18)&[MHH96]J120226-012948\nl
12018+1941&0.168&&0.168, 0.169&(4), (19)\nl
12032+1707&0.217&&0.168&(19)\nl
12072$-$0444&0.129&&0.128&(14)\nl
12112+0305&0.073&&0.073, 0.072&(13), (18)\nl
12127$-$1412&0.133&&\nodata&\nodata\nl
12265+0219&\nodata&&0.159, 0.158&(15), (18)&3C273, PG1226+023, H1226+023\nl
12359$-$0725&0.138&&\nodata&\nodata\nl
\nl
12447+3721&0.158&&\nodata&\nodata\nl
12540+5708&0.042&&0.042&(3)&UGC08058, Mrk231, VIIZw490\nl
13106$-$0922&0.174&&0.174&(19)\nl
13218+0552&0.205&&0.190, 0.205&(9), (19)&[HB89]1321+058, H1321+058\nl
13305$-$1739&0.148&&0.148&(1)\nl
13335$-$2612&0.125&&0.125&(1)\nl
13342+3932&0.179&&0.179&(17)\nl
13428+5608&0.037&&0.037&(13)&UGC8696, Mrk273, IZw71, VV851\nl
13443+0802&0.135&&0.135&(19)\nl
13451+1232&0.122&&0.122&(6)&4C12.50\nl
\nl
13454$-$2956&0.129&&0.129, 0.076&(1), (18)\nl
13469+5833&0.158&&0.158&(19)\nl
13509+0442&0.136&&0.189&(19)\nl
13539+2920&0.108&&0.108&(18)\nl
14053$-$1958&0.161&&0.162&(1)\nl
14060+2919&0.117&&0.117&(17)\nl
14070+0525&0.265&&0.265&(19)\nl
14121$-$0126&0.151&&0.150&(19)\nl
14197+0813&0.131&&\nodata&\nodata\nl
14202+2615&0.159&&0.158&(17)\nl
\nl
14252$-$1550&0.149&&0.150&(19)\nl
14348$-$1447&0.083&&0.082&(13)\nl
14394+5332&0.105&&0.105&(17)\nl
14485$-$2434&0.148&&\nodata\nl
15001+1433&0.162&&0.163&(18)\nl
15043+5754&0.151&&\nodata&\nodata\nl
15130$-$1958&0.109&&0.109&(18)\nl
15206+3342&0.125&&0.125, 0.124&(14), (19)\nl
15225+2350&0.139&&0.138&(19)\nl
15327+2340&0.018&&0.018&(12)&UGC09913, IC4553/4, VV540, Arp220\nl
\nl
15462$-$0450&0.100&&0.101&(18)\nl
16090$-$0139&0.134&&0.134&(18)\nl
16156+0146&0.132&&\nodata&\nodata\nl
16300+1558&0.242&&0.242&(19)\nl
16333+4630&0.191&&0.191&(19)\nl
16468+5200&0.150&&\nodata&\nodata\nl
16474+3430&0.111&&0.111&(18)\nl
16487+5447&0.104&&0.104&(18)\nl
17028+5817&0.106&&0.106&(18)\nl
17044+6720&0.135&&0.135&(19)\nl
\nl
17068+4027&0.179&&0.072&(19)\nl
17179+5444&0.147&&0.147&(19)&TXS1717+547\nl
20414$-$1651&0.086&&0.087&(18)\nl
21208$-$0519&0.130&&\nodata&\nodata\nl
21219$-$1757&0.112&&0.113&(8)&[HB89]2121-179\nl
21329$-$2346&0.125&&0.125&(19)\nl
21477+0502&0.171&&\nodata&\nodata\nl
22088$-$1831&0.170&&0.170&(19)\nl
22206$-$2715&0.132&&0.131&(19)\nl
22491$-$1808&0.076&&0.076, 0.078&(16), (18)\nl
\nl
22541+0833&0.166&&0.166&(19)\nl
23060+0505&0.184&&0.174&(7)\nl
23129+2548&0.179&&\nodata&\nodata\nl
23233+2817&0.114&&\nodata&\nodata\nl
23234+0946&0.128&&0.128&(19)\nl
23327+2913&0.107&&0.107&(18)\nl
23389+0300&0.145&&0.145&(19)&4C03.60\nl
23498+2423&0.212&&\nodata&\nodata\nl
\tablerefs{Redshifts:  1 - Allen et al. (1991), 2 - Arakelyan et al. (1971), 
3 - Armus, Heckman \& Miley (1989), 4 - Baan (1989), 5 - Bottinelli et al. (1987)
6 - Grandi (1977), 7 - Hill, Wynn-Williams \& Becklin (1987), 8 - Hill et al. (1988), 
9 - Low et al. (1988), 10 - MacKenty \& Stockton (1984), 
11 - Melnick \& Mirabel (1990), 12 - Mirabel (1982), 13 - Sanders et al. (1988a), 
14 - Sanders et al. (1988b), 15 - Schmidt (1963), 16 - Soifer et al. (1987), 
17 - Strauss \& Huchra (1988), 18 - Strauss et al. (1992), 19 - Fisher et al. (1995)}
\enddata
\end{deluxetable}

\clearpage

\begin{deluxetable}{cll}
\tablenum{3}
\tablewidth{4.5in}
\tablecaption{Luminosity Function}
\tablehead{
\nl
\colhead{log $L_{\rm ir}$} &
\multicolumn{1}{c}{$\Phi$} &
\multicolumn{1}{c}{$\langle V / V_{\rm m} \rangle$\tablenotemark{a}} \nl
\colhead{$\lbrack L_\odot \rbrack$} &
\multicolumn{1}{c}{$\lbrack {\rm Mpc}^{-3} \left({\rm Mag}/2 \right)^{-1} \rbrack$} &
\multicolumn{1}{c}{} \nl
}
\startdata
12.0--12.2&$1.3 \pm 0.2 \times 10^{-7}$&$0.55 \pm 0.03$ (63) \nl
12.2--12.4&$3.9 \pm 0.7 \times 10^{-8}$&$0.55 \pm 0.06$ (35) \nl
12.4--12.6&$9.6 \pm 2.6 \times 10^{-9}$&$0.50 \pm 0.08$ (14) \nl
12.6--12.8&$2.2 \pm 1.0 \times 10^{-9}$&$0.61 \pm 0.15$ (5) \nl
12.8--13.0&$1.7 \pm 1.7 \times 10^{-10}$&$0.40 \pm 0.29$ (1) \nl
\tablenotetext{a}{Number in parenthesis represents number of galaxies used for statistics.}
\enddata
\end{deluxetable}